\documentclass[12pt]{article}
\usepackage{graphicx}
\begin{document}

\begin{center}

{\Large \bf Lorentz Group applicable to Finite Crystals}

\vspace{7mm}

Sibel Ba{\c s}kal \footnote{electronic address:
baskal@newton.physics.metu.edu.tr}\\

Department of Physics, Middle East Technical University,
06531 Ankara, Turkey

\vspace{5mm}

Elena Georgieva\footnote{electronic address:
egeorgie@pop500.gsfc.nasa.gov}\\
Science Systems and Applications, Inc., Lanham, MD 20771,
\\ and \\
National Aeronautics and Space Administration, \\ Goddard
Space Flight Center,  Laser and Electro-Optics Branch,
Code 554, Greenbelt, Maryland 20771

\vspace{5mm}

Y. S. Kim\footnote{electronic address: yskim@physics.umd.edu}
\\
Department of Physics, University of Maryland,\\
College Park, Maryland 20742

\vspace{5mm}

\end{center}

\begin{abstract}
The two-by-two scattering matrix for one-dimensional
scattering processes is a three-parameter $Sp(2)$ matrix or
its unitary equivalent.  For one-dimensional crystals, it
would be repeated applications of this matrix.  The problem
is how to calculate $N$ repeated multiplications of this
matrix.  It is shown that the original $Sp(2)$ matrix can be
written as a similarity transformation of Wigner's little
group matrix which can be diagonalized.  It is then possible
to calculate the repeated applications of the original
$Sp(2)$ matrix.
\end{abstract}

\vspace{5mm}

Besides the three-dimensional rotation matrix, one of the
most commonly used matrices in physics is the two-by-two
$Sp(2)$ matrix.  First of all, it is isomorphic to the
Lorentz group applicable to two space dimensions and one time
variable.  It is the fundamental building block for linear
canonical transformations for phase-space approach to both
classical and quantum mechanics~\cite{knp91}.  The same is
true for classical and quantum optics.

In addition, the one-dimensional scattering or beam-transfer
matrix takes the form of a $Sp(2)$ matrix or its unitary
equivalent.  Indeed, this $Sp(2)$ matrix takes a simple
mathematical form.  It is real and unimodular, and therefore
has three independent parameters.  The question then arises
as how to handle the calculation when this matrix is applied
repeatedly for finite one-dimensional crystals, including
periodic potentials~\cite{sprung93,griffith01} and multilayer
optics~\cite{gk03,sanch05}.

In spite of its fundamental role in those many branches of
physics, its mathematics is not yet thoroughly understood.
First of all, there does not seem to exist an established
procedure for diagonalizing this matrix.  It is not difficult
to get its eigenvalues, but the matrix cannot be diagonalized
by a rotation.  Is it then possible to construct a similarity
transformation which brings it to a diagonal form?  If it is
possible, the most immediate application is the scattering of
beams in the above-mentioned one-dimensional crystals which can
mathematically be formulated through beam transfer matrices.

In this report, we show that the most general form of the
$Sp(2)$ matrix can be written as a similarity transformation
of Winger's little group matrix~\cite{wig39,knp86}.  This
Wigner matrix takes three different forms depending on the
parameters of the original $Sp(2)$ matrix.  However, these
forms are either  diagonal or can be diagonalized by a
rotation.

In most of the physical applications, the one-cycle beam
transfer matrix can be written as
\begin{equation}\label{eq01}
R_1 B R_2
\end{equation}
with
\begin{eqnarray}\label{eq01a}
&{}& R_i = \pmatrix{\cos(\theta_i/2)  &  -\sin(\theta_i/2)
\cr   \sin(\theta_i/2)  &  \cos(\theta_i/2)} , \nonumber
\\[2ex]
&{}& B = \pmatrix{e^\lambda & 0 \cr 0 & e^{-\lambda}} .
\end{eqnarray}
This is known as the Bargmann decomposition of the $Sp(2)$
matrix.  The original three-parameter matrix is written as
a product of three one-parameter matrices.

We can write the expression of Eq.(\ref{eq01}) as
\begin{equation}\label{eq02}
R_1 B R_2  = (D R) B \left(R D^{-1}\right) ,
\end{equation}
where
\begin{equation}\label{r01}
R = \sqrt{R_1 R_2} = \pmatrix{\cos(\theta/2)  &
-\sin(\theta/2) \cr \sin(\theta/2)  &  \cos(\theta/2)} ,
\end{equation}
with
$$
\theta = \frac{\theta_1 + \theta_2}{2} ,
$$
and
\begin{equation}\label{d01}
D = \sqrt{R_1 \, R_2\,^{-1}} = \pmatrix{\cos(\delta/2)  &
-\sin(\delta/2) \cr \sin(\delta/2)  &  \cos(\delta/2)} ,
\end{equation}
with
$$
\delta = \frac{\theta_1 - \theta_2}{2},
$$
and
\begin{equation}
R_1 R_2 = R_2 R_1 = R^2, \quad R_1 = D R, \quad
R_2 = R D^{-1} .
\end{equation}
We can thus write the starting matrix of Eq.(\ref{eq01}) as
\begin{equation}~\label{eq05}
R_1 B R_2 = D (R B R) D^{-1} .
\end{equation}
This is a similarity transformation of $(R B R)$ with
respect to $D$.

In this report, we are interested in diagonalizing the
expression of Eq.(\ref{eq05}), and thus calculating
$\left(R_1 B R_2\right)^N$.  For this purpose, we point
out that $(R B R)$ can be written as a similarity
transformation~\cite{hk88,gk03}
\begin{equation}
R B R = S W S^{-1} ,
\end{equation}
where $S$ is in the form of
\begin{equation}
S = \pmatrix{e^{\eta/2} & 0 \cr 0 & e^{-\eta/2} } ,
\end{equation}
and $W$ is Wigner's little-group matrix which takes one
of the following three forms~\cite{gk03}.
\begin{eqnarray}\label{wlg01}
&{}& R(\phi) = \pmatrix{\cos(\phi/2)  & -\sin(\phi/2) \cr
         \sin(\phi/2)  & \cos(\phi/2) } , \nonumber \\[2ex]
&{}& X(\chi) = \pmatrix{\cosh(\chi/2) & \sinh(\chi/2)  \cr
                     \sinh(\chi/2) & \cosh
(\chi/2)} ,\nonumber \\[2ex]
&{}& E(\gamma) = \pmatrix{1 & \gamma \cr 0 & 1},
\quad \mbox{or} \quad
           E(\gamma) = \pmatrix{1 & 0 \cr -\gamma & 1} .
\end{eqnarray}

We can then write
\begin{equation}\label{eq10}
\left(R_1 B R_2\right) = D \left(S W S^{-1}\right) D^{-1}
    = (DS) W (DS)^{-1} .
\end{equation}
Indeed, the most general form of the $Sp(2)$ matrix can be
written as a similarity transformation of the Wigner's
little-group matrix.

The immediate application of this formula is to calculate
$\left(R_1 B R_2 \right)^N$.  Although two-by-two matrices
are simple, it is not trivial to calculate this expression
for a large number of N.  On the other hand,
from Eq.(\ref{eq10}), it is straightforward to write
\begin{equation}
\left(R_1 B R_2 \right)^N = (DS) W^N (DS)^{-1} ,
\end{equation}
while it is a simple matter to calculate $W^N$ from the
expression given in
Eq.(\ref{wlg01}).  The expression for $W^N$ becomes $R
(N\phi), X(N\chi)$
or $ E(N\gamma)$.

The remaining problem is to relate the parameters $\eta$
of the $S$ matrix and $\phi, \chi$ or $\gamma$ of the $W$
matrix from $\lambda$ of $B$ in Eq.(\ref{eq01a}) and
$\theta$ of $R$ in Eq.(\ref{r01}).  The calculation is
straightforward~\cite{gk03,bk03}.  When
$(\cosh\lambda \sin\theta - \sinh\lambda)$ is positive,
the relation becomes $R(\theta) B R(\theta)  =
S R(\phi) S^{-1}$, and the result is
\begin{eqnarray}\label{eqn22}
&{}& \cos(\phi/2) = \cosh\lambda \cos\theta, \nonumber
\\[3ex]
&{}& e^{2\eta} = {\cosh\lambda \sin\theta + \sinh\lambda
\over
     \cosh\lambda \sin\theta - \sinh\lambda} .
\end{eqnarray}

If $(\cosh\lambda \sin\theta - \sinh\lambda)$ is negative,
we should use $X(\chi)$ as the little-group matrix, and
write $R(\theta) B R(\theta) = S X(\chi) S^{-1}$. The result
is
\begin{eqnarray}
\cosh(\chi/2) = \cosh\lambda \cos\theta, \nonumber \\[2ex]
e^{2\eta} = {\cosh\lambda \sin\theta + \sinh\lambda  \over
 \sinh\lambda - \cosh\lambda \sin\theta } .
\end{eqnarray}

When $(\cosh\lambda \sin\theta - \sinh\lambda)$ goes through
zero while it makes a transition from a small negative to
positive number, is negative, we should use $R(\theta) B
R(\theta) = S E(\gamma) S^{-1}$, the result is
\begin{equation}
e^{\eta}\sin(\phi/2) = \gamma.
\end{equation}
In this case, $e^{\eta}$ becomes very large, and $\phi$ has
to be very small for $\gamma$ to be finite~\cite{gk03,bk03}.

In this report, we started with the most general form of
the $Sp(2)$ matrix and its Bargmann decomposition.  We then
showed that it can also be written as a similarity
transformation of Wigner's little group matrix, which can be
diagonalised by a rotation.  The immediate application of
this similarity transformation is in the one-dimensional
crystals requiring repeated applications of the $Sp(2)$
matrix.

Although the $Sp(2)$ matrix is mathematically convenient,
its physical application is often made through its unitary
equivalent~\cite{gk01,bk02}.  Transformations among those
equivalent representations are also mathematically
challenging problems.

\end{document}